\documentclass[final,5p,times,twocolumn]{elsarticle}

\usepackage{amssymb}
\usepackage{color} 

\journal{Journal of Magnetism and Magnetic Materials}

\begin{document}

\begin{frontmatter}



\title{Frustrated diamond-chain quantum $XXZ$ Heisenberg antiferromagnet in 
a magnetic field}


\author{Johannes Richter$^a$, Olesia Krupnitska$^b$, Taras Krokhmalskii$^{b,c}$, Oleg Derzhko$^{a,b,c,d}$}

\address{$^a$Institut f\"{u}r theoretische Physik,
          Otto-von-Guericke-Universit\"{a}t Magdeburg, P.O. Box 4120, D-39016 Magdeburg, Germany\\
$^b$Institute for Condensed Matter Physics,
          National Academy of Sciences of Ukraine,  
          1 Svientsitskii Street, L'viv-11, 79011, Ukraine\\
$^c$Department for Theoretical Physics, 
          Ivan Franko National University of L'viv,
          12 Drahomanov Street, L'viv-5, 79005, Ukraine\\
$^d$Abdus Salam International Centre for Theoretical Physics,
          Strada Costiera 11, I-34151 Trieste, Italy}

\begin{abstract}
We consider the antiferromagnetic spin-1/2 $XXZ$ Heisenberg model on a
frustrated diamond-chain lattice
in a $z$- or $x$-aligned external magnetic field.
We use the strong-coupling approach to elaborate an effective description in
the low-temperature strong-field regime.
The obtained effective models are spin-1/2 $XY$ chains which are exactly
solvable through the Jordan-Wigner fermionization.
We perform exact-diagonalization studies of the magnetization curves to test
the quality of the effective description.
The results may have relevance for the description of the azurite spin-chain
compound.
\end{abstract}

\begin{keyword}
quantum $XXZ$ Heisenberg antiferromagnet \sep
          geometrical frustration \sep
          magnetization curve
\PACS 75.10.Jm




\end{keyword}

\end{frontmatter}



\section{Introduction}
\label{sec1}

Many frustrated magnetic compounds with strong quantum
fluctuations have been synthesized and measured during last years.
A challenging target is to provide a theoretical description of the observable properties.
An important class of frustrated quantum Heisenberg antiferromagnets,
for which the properties can be described in great detail in the low-temperature strong-field regime,
is the class of the so-called localized-magnon systems.\citep{locmag1,locmag2,locmag3}
These localized-magnon systems have a completely dispersionless (flat) lowest-energy one-magnon band 
that offers the  possibility to consider one-magnon states which are localized within a small part of the lattice (trap).
As a result,
many-magnon ground states can be constructed, their degeneracy can be calculated with the help of an auxiliary classical lattice-gas model,
and the low-temperature strong-field thermodynamics of the quantum systems at hand can be elaborated 
using the methods of classical statistical mechanics.\citep{locmag2,locmag3}

From the aspect of solid-state physics,
it is important to notice the following.
The lowest-energy magnon band is strictly flat only for special relations between exchange couplings 
that corresponds to one particular point in the parameter space.
One cannot expect that a real-life system would reach exactly this flat-band point.
However, a certain magnetic compound may be quite close to it.
A fascinating example is the natural mineral azurite Cu$_3$(CO$_3$)$_2$(OH)$_2$,\citep{kikuchi,jeschke}
which is known as a realization of the frustrated diamond spin chain close to the flat-band point.\citep{jeschke,jpcm}
In our recent study\citep{prb-fnt} 
we have considered three localized-magnon systems 
(the diamond and dimer-plaquette chains and the two-dimensional square-kagome lattice) 
which belong to the monomer universality class\citep{locmag3}
with small deviations from the ideal flat-band geometry
and have elaborated effective low-energy theories
which provide good description of the initial frustrated quantum isotropic (i.e., $XXX$) Heisenberg antiferromagnets in the low-temperature strong-field regime.

In the present paper we make one step further and consider an anisotropic $XXZ$ Heisenberg (antiferromagnetic) interaction 
to examine the dependence of some observables at low temperatures on the orientation of the applied (strong) magnetic field.
Note that for the isotropic $XXX$ Heisenberg interaction the orientation of the applied magnetic field is irrelevant.

From the experimental point of view,
we know that the low-temperature magnetization curves for azurite
with the field applied along chain direction and perpendicularly to chain direction 
are different\citep{kikuchi,kikuchi-2} 
that indicates an anisotropic exchange interaction.
Even stronger anisotropy effects have been recently observed for the spin dimer magnet Ba$_2$CoSi$_2$O$_6$Cl$_2$
-- a recently synthesized compound which should also exhibit localized-magnon physics.\citep{tanaka}
In this compound the antiferromagnetic exchange interactions between the Co$^{2+}$ spin-1/2 ions are
$XY$-like
and the exchange-interaction network of the Co$^{2+}$ sites is described by a frustrated bilayer lattice.\citep{frustrated_bilayer,fb}

In what follows we investigate  the frustrated diamond-chain lattice as a paradigmatic system to study localized-magnon effects.
While the case of the $z$-aligned magnetic field does not require a new framework for getting effective models
in comparison with the previously reported scheme,\citep{prb-fnt}
the case of a magnetic field aligned in $x$ or $y$ direction is different, 
since the Zeeman term does not commute with the anisotropic $XXZ$ Hamiltonian any more. 
We consider a field aligned along the $x$-axis 
(trivially, for symmetry reasons a $y$-aligned field leads to the same results). 
Furthermore, motivated by the experimental data for Ba$_2$CoSi$_2$O$_6$Cl$_2$
we focus on antiferromagnetic $XXZ$ interactions with easy-plane anisotropy. 
To obtain effective Hamiltonians we use the strong-coupling approach.\citep{andreas-andreas,jpcm}
After deriving the effective Hamiltonians we test their quality by comparisons to exact-diagonalization data.
Using exact results for the effective models 
we can provide theoretical predictions for the low-temperature magnetization curves in $z$- or $x$-aligned (strong) magnetic field 
as well as for other thermodynamic quantities in this regime
in the thermodynamic limit.

\section{The model and effective theories}
\label{sec2}

\begin{figure}[h]
\begin{center}
\includegraphics[clip=on,width=60mm,angle=0]{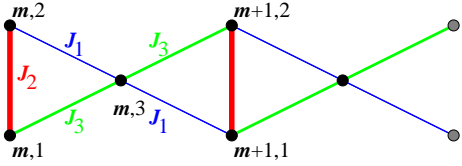}
\vspace{1mm}
\caption
{(Color online) The distorted frustrated diamond-chain lattice considered in this paper.
The lattice sites are labeled by a pair of indeces, 
where the first number enumerates the unit cells
($m=1,\ldots,{\cal{N}}$, ${\cal{N}}=N/3$)
and the second one enumerates the position of the site within the unit cell.
The localized-magnon picture holds if $J_1=J_3<J_2/2$, see Ref.~\citep{locmag3}.}
\label{fig01}
\end{center}
\end{figure}

To be specific, we consider the spin-1/2 Hamiltonian
\begin{eqnarray}
\label{00}
H=\sum_{(ij)}J_{ij}\left(s_i^xs_j^x+s_i^ys_j^y+\Delta s_i^zs_j^z\right)-{\bf{h}}\cdot {\bf{S}}
\end{eqnarray}
on a $N$-site frustrated diamond-chain lattice shown in Fig.~\ref{fig01}.
Here the sum runs over all neighboring bonds, 
$J_{ij}>0$,
$0\le \Delta \le 1$,
and ${\bf{S}}=\sum_i {\bf{s}}_i$.
The flat-band case is realized for $J_1=J_3<J_2/2$, i.e., $J_2$ is the dominating interaction.
We consider two field orientations,
${\bf{h}}=(0,0,h)$ ($h^z$ field)
and
${\bf{h}}=(h,0,0)$ ($h^x$ field),
and assume the applied field $h>0$ to be strong.
Our aim is to find an effective low-energy theory by exploiting the strong-coupling approach.\citep{andreas-andreas}
In case of the $x$-aligned field 
we first perform a $\pi/2$ rotation around the $y$ axes in the spin space.
As a result we get a Hamiltonian again with anisotropic exchange interaction but now in a $z$-aligned field   
\begin{eqnarray}
\label{03}
H=\sum_{(ij)}J_{ij}\left[{\bf{s}}_i\cdot
{\bf{s}}_j+\left(\Delta-1\right)s_i^xs_j^x\right]-h S^z.
\end{eqnarray}

Having in mind the dominance of the vertical dimer bond $J_2$ 
we apply on both models (\ref{00}) with ${\bf{h}}=(0,0,h)$ and (\ref{03}) the strong-coupling approach. 
There are some common features of this approach valid for both cases.
First  
we use as a starting point a collection of ${\cal{N}}=N/3$ vertical dimers ($J_2$ bonds)  
and the ${\cal{N}}$ isolated sites labeled by $m,3$ 
(see Fig.~\ref{fig01})
at the ``bare'' saturation field $h_0$. 
We consider this part of the Hamiltonian $H$ as the main
Hamiltonian
$H_{\rm{main}}$.
Hence the spins at the sites $m,3$ are in the spin-up state.
The spin problem on the dimer is solved analytically
and we find two relevant low-energy states $\vert u\rangle$ and $\vert d\rangle$,
where the energies of these states, $\epsilon_u$ and $\epsilon_d$, coincide if $h=h_0$.
This produces a $2^{\cal{N}}$-fold degeneracy of the ground state $\vert\varphi_0\rangle$ of $H_{\rm{main}}$ at $h=h_0$.
Then we switch on the perturbation $V=H-H_{\rm{main}}$ which contains the interaction terms $J_1$ and $J_3$ 
(see Fig.~\ref{fig01}).
Introducing the projector onto a model space $P=\vert\varphi_0\rangle\langle\varphi_0\vert$,
for the effective Hamiltonian $H_{\rm{eff}}$, 
which acts in the model space only but gives the exact ground-state energy of $H$,
we get perturbatively\citep{fulde}
\begin{eqnarray}
\label{01}
H_{\rm{eff}}=PHP+PV\sum_{\alpha\ne 0}\frac{\vert\varphi_{\alpha}\rangle\langle\varphi_{\alpha}\vert}{\varepsilon_0-\varepsilon_{\alpha}}VP+\ldots
\; .
\end{eqnarray}
Here $\vert\varphi_\alpha\rangle$ ($\alpha\ne 0$) are excited states of $H_{\rm{main}}$.
Finally we use the (pseudo)spin-1/2 operators
$T^z=(\vert u\rangle\langle u\vert -\vert d\rangle\langle d\vert)/2$,
$T^+=\vert u\rangle\langle d\vert$,
and
$T^-=\vert d\rangle\langle u\vert$
to write the effective Hamiltonians in terms of spin
operators thus recognizing that the effective Hamiltonians correspond to
some well-known quantum spin
models.

\subsection{The case ${\bf{h}}=(0,0,h)$}

We begin with the case of the $z$-aligned field, see Hamiltonian (\ref{00}).
The calculations are quite similar to the ones explained previously\citep{prb-fnt}
and the final result for effective Hamiltonian (\ref{01}) reads:
\begin{eqnarray}
\label{02}
H_{\rm{eff}}=\sum_{m}\left[{\sf{C}}-{\sf{h}}T_m^z + {\sf{J}}\left(T_m^xT_{m+1}^x+T_m^yT_{m+1}^y\right)\right],
\nonumber\\
{\sf{C}}=-h-\frac{J_2}{4}+\Delta\frac{J}{2}-\frac{(J_3-J_1)^2}{4(1+\Delta)J_2},
\;\;\;
J=\frac{J_3+J_1}{2},
\nonumber\\
{\sf{h}}=h-h_1-\frac{(J_3-J_1)^2}{2(1+\Delta)J_2},
\;\;\;
h_1=\frac{1+\Delta}{2}J_2+\Delta J,
\nonumber\\
{\sf{J}}=\frac{(J_3-J_1)^2}{2(1+\Delta)J_2}. \qquad \qquad \qquad \qquad 
\end{eqnarray}
In the limit $\Delta=1$ this result reproduces the effective
Hamiltonian obtained
previously in Refs.
\cite{andreas-andreas,jpcm,prb-fnt}.
For $0\le \Delta <1$ there are only quantitative changes in the parameters of the effective Hamiltonian.

\subsection{The case ${\bf{h}}=(h,0,0)$}

The case of the $x$-aligned field [Hamiltonian (\ref{03}) in the rotated coordinate frame] requires more explanations.
From Eq.~(\ref{03}) it is obvious that the Zeeman term does not commute with the Hamiltonian.
The two relevant low-energy states of each $J_2$-bond are:
\begin{eqnarray}
\label{04}
\vert u\rangle &=& a\vert\uparrow_1\uparrow_2\rangle+b\vert\downarrow_1\downarrow_2\rangle,
\nonumber\\
a &=& \frac{1}{C}\left[h+\sqrt{\frac{(1-\Delta)^2}{16}J_2^2+h^2}\right],
\;\;\;
b=\frac{1}{C}\frac{1-\Delta}{4}J_2,
\nonumber\\
C \hspace{-5pt} &=&  \hspace{-5pt} \sqrt{2}\sqrt{\frac{(1-\Delta)^2}{16}J_2^2+h\sqrt{\frac{(1-\Delta)^2}{16}J_2^2+h^2}+h^2}
\end{eqnarray}
with the energy $\epsilon_u=\big [J_2-\sqrt{(1-\Delta)^2J_2^2+16h^2}\big ]/4
\;$
and
\begin{eqnarray}
\label{05}
\vert d\rangle=\frac{1}{\sqrt{2}}\left(\vert\uparrow_1\downarrow_2\rangle -\vert\downarrow_1\uparrow_2\rangle\right)
\end{eqnarray}
with the energy $\epsilon_d=-(2+\Delta)J_2/4$.
Furthermore, 
the ``bare'' saturation field is
$h_0=\sqrt{(1+\Delta)/2}\;J_2$.
For the first term in the r.h.s. of Eq.~(\ref{01}) we get by straightforward calculations
\begin{eqnarray}
\label{06}
PHP=\sum_m\left[-\frac{h}{2}-\frac{2+\Delta}{4}J_2  \hspace{2cm}
\right.
\nonumber\\
\left.
-\left(h-h_0-J\right)\left(a^2-b^2\right)\left(\frac{1}{2}+T_m^z\right)
\right],
\end{eqnarray}
where again $J=(J_3+J_1)/2$.
There are two sets of excited states which contribute to the second term in the r.h.s. of Eq.~(\ref{01}).
The first one consists of ${\cal{N}}2^{\cal{N}}$ states
which differ from $\vert\varphi_0\rangle$ by one flipped spin at the site $m,3$
(see Fig.~\ref{fig01}).
Their energy is $\varepsilon_{\alpha_1}=\varepsilon_0+h_0$.
Furthermore, we find
\begin{eqnarray}
\label{07}
&& PV\sum_{\alpha_1\ne 0}\frac{\vert\varphi_{\alpha_1}\rangle\langle\varphi_{\alpha_1}\vert}{\varepsilon_0-\varepsilon_{\alpha_1}}VP
\nonumber\\
&& =\sum_{m}\left({\sf{C}}_1-{\sf{h}}_1T_m^z+{\sf{J}}^xT_m^xT_{m+1}^x+{\sf{J}}^yT_m^yT_{m+1}^y\right),
\nonumber\\
&& {\sf{C}}_1=-\frac{(J_3-J_1)^2}{16h_0}\left[1+2ab(1-\Delta^2)+\Delta^2\right],
\nonumber\\
&& {\sf{h}}_1=-\frac{(J_3-J_1)^2}{4h_0}\left(a^2-b^2\right)\Delta,
\nonumber\\
&& {\sf{J}}^x=\frac{(J_3-J_1)^2}{4h_0}\left(a-b\right)^2\Delta^2,
\nonumber\\
&& {\sf{J}}^y=\frac{(J_3-J_1)^2}{4h_0}\left(a+b\right)^2.
\end{eqnarray}
The second set of excited states consists of ${\cal{N}}2^{{\cal{N}}-1}$ states
which differ from the ground state $\vert\varphi_0\rangle$ by
a dimer state
$-b\vert\uparrow_1\uparrow_2\rangle+a\vert\downarrow_1\downarrow_2\rangle$
(with energy $[J_2+\sqrt{(1-\Delta)^2J_2^2+16h^2}]/4$) in a certain cell $m$
(see also Fig.~\ref{fig01}).
The energy of these excited states is $\varepsilon_{\alpha_2}=\varepsilon_0+(3+\Delta)J_2/2$.
Furthermore, we find
\begin{eqnarray}
\label{08}
PV\sum_{\alpha_2\ne 0}\frac{\vert\varphi_{\alpha_2}\rangle\langle\varphi_{\alpha_2}\vert}{\varepsilon_0-\varepsilon_{\alpha_2}}VP
\hspace{4cm}
\nonumber\\
=-\sum_{m}\frac{8\left(h-h_0-J\right)^2}{(3+\Delta)J_2}a^2b^2\left(\frac{1}{2}+T_m^z\right).
\end{eqnarray}
Combining Eqs.~(\ref{06}), (\ref{07}), and (\ref{08}) we arrive at
\begin{eqnarray}
\label{09}
H_{\rm{eff}}&=&\sum_{m}\left({\sf{C}}-{\sf{h}}T_m^z+{\sf{J}}^xT_m^xT_{m+1}^x+{\sf{J}}^yT_m^yT_{m+1}^y\right),
\nonumber\\
{\sf{C}}
&=&
-\frac{h}{2}-\frac{2+\Delta}{4}J_2
-\frac{1}{2}\left(h-h_0-J\right)\left(a^2-b^2\right)
\nonumber\\
&&+{\sf{C}}_1
-\frac{4\left(h-h_0-J\right)^2}{(3+\Delta)J_2}a^2b^2,
\nonumber\\
{\sf{h}}
&=&
\left(h-h_0-J\right)\left(a^2-b^2\right)
\nonumber\\
&& +{\sf{h}}_1
+\frac{8\left(h-h_0-J\right)^2}{(3+\Delta)J_2}a^2b^2,
\end{eqnarray}
where ${\sf{J}}^x$, ${\sf{J}}^y$, ${\sf{C}}_1$, and ${\sf{h}}_1$
are given in Eq.~(\ref{07}).

Let us briefly discuss the obtained effective Hamiltonian (\ref{09}).
In the limit $\Delta=1$ it reproduces the strong-coupling findings reported
previously.\citep{andreas-andreas,jpcm,prb-fnt}
Clearly, in this limit both effective theories, given in Eq.~(\ref{02}) and in Eq.~(\ref{09}), coincide,
i.e., ${\sf{J}}^x={\sf{J}}^y={\sf{J}}$.
If $\Delta$ becomes smaller than one, ${\sf{J}}^x$ becomes smaller than ${\sf{J}}^y$. 
Finally, ${\sf{J}}^x$ vanishes in the $XY$-limit ($\Delta=0$).

\subsection{Summary of analytical findings}

To summarize,
we have found that the frustrated diamond-chain quantum $XXZ$ Heisenberg antiferromagnet
[see Eq.~(\ref{00}) and Fig.~\ref{fig01}]
in a strong magnetic field ${\bf{h}}=(0,0,h)$ or ${\bf{h}}=(h,0,0)$ 
is described by the effective low-energy theory given in Eq.~(\ref{02}) or Eq.~(\ref{09}).
The emergent effective spin-1/2 chains,
Eqs.~(\ref{02}) and (\ref{09}),
are exactly solvable by Jordan-Wigner fermionization,
see, e.g., Ref.~\citep{fermionization}.
Therefore the free energy (per cell) of the initial frustrated quantum spin model in the low-temperature strong-field regime for ${\cal{N}}\to\infty$ is given by
\begin{eqnarray}
\label{10}
f(T,h)={\sf{C}}-\frac{T}{2\pi}\int_{-\pi}^{\pi}{\rm{d}}\kappa \ln\left(2\cosh\frac{\Lambda_\kappa}{2T}\right)
\end{eqnarray}
with 
\begin{eqnarray}
\label{11}
\Lambda_\kappa=\sqrt{\left(-{\sf{h}}+\frac{{\sf{J}}^x+{\sf{J}}^y}{2}\cos\kappa\right)^2+\left(\frac{{\sf{J}}^x-{\sf{J}}^y}{2}\sin\kappa\right)^2}.
\end{eqnarray}
For the $z$-aligned or $x$-aligned magnetic field the effective model parameters in Eqs.~(\ref{10}) and (\ref{11}) are given in Eq.~(\ref{02}) or in Eq.~(\ref{09}).
It is worth noting
that in case of ideal flat-band geometry ($J_1=J_3$) 
the interaction terms in the effective models vanish: ${\sf{J}}={\sf{J}}^x={\sf{J}}^y=0$.
As a result, 
Eqs.~(\ref{10}) and (\ref{11}) simplify and in the case of the $z$-aligned field they reproduce the localized-magnon theory of Ref.~\citep{locmag3}.

Knowing the free energy (\ref{10}) one can easily obtain all thermodynamic quantities.
For example,
the magnetization per cell is given by ${\cal{M}}(T,h)=-\partial f(T,h)/\partial h$
[the magnetization per site is three times smaller,
$m(T,h)={\cal{M}}(T,h)/3$].

\section{Comparison to exact-diagonalization data}
\label{sec3}

\begin{figure}
\begin{center}
\includegraphics[clip=on,width=75mm,angle=0]{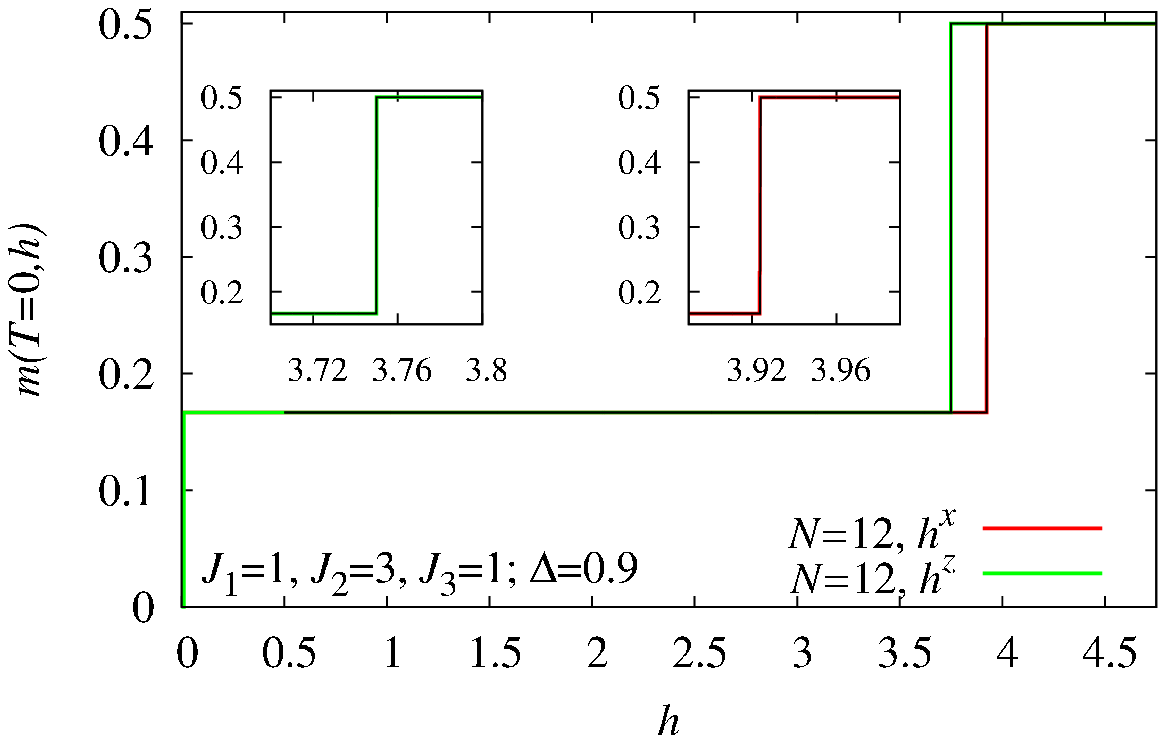}\\
\vspace{3mm}
\includegraphics[clip=on,width=75mm,angle=0]{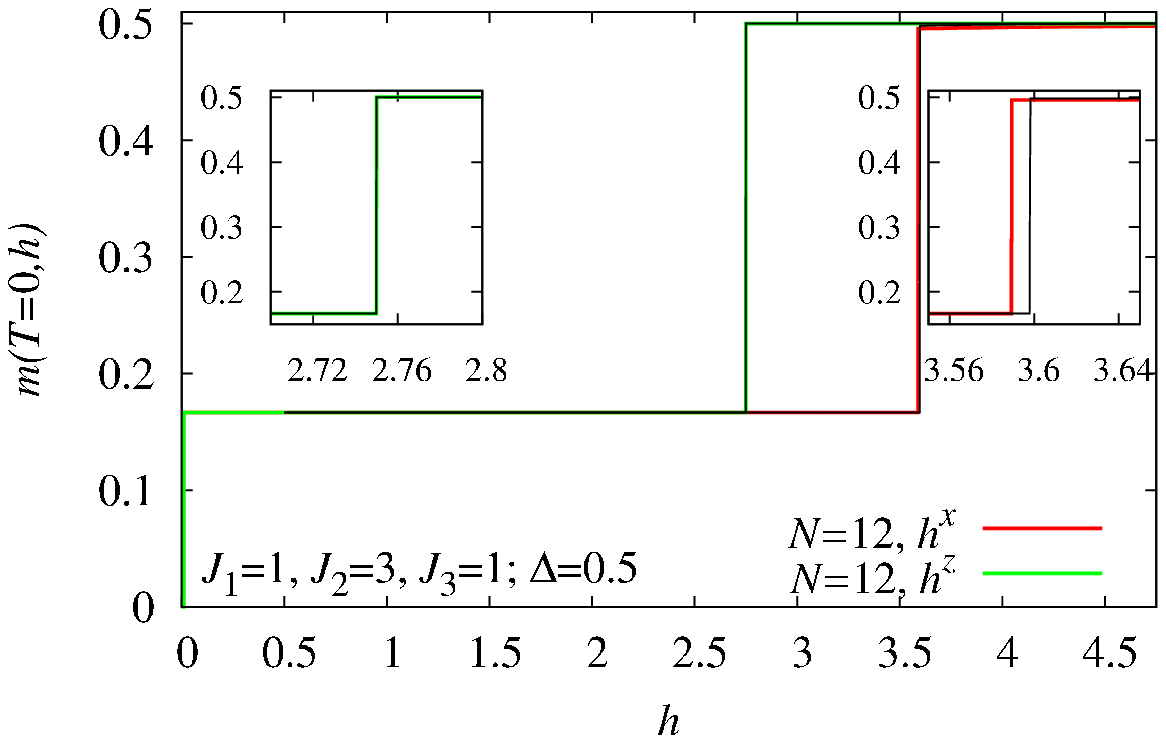}\\
\vspace{3mm}
\includegraphics[clip=on,width=75mm,angle=0]{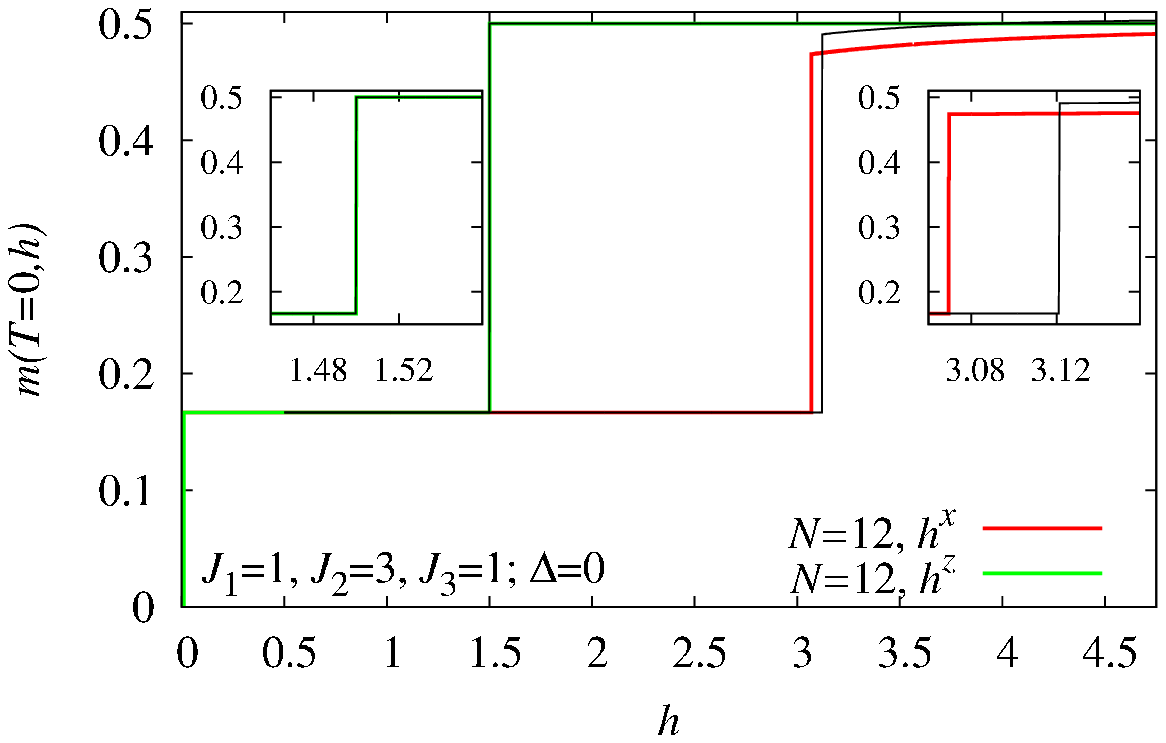}
\caption
{(Color online) Ground-state magnetization curves $m(T=0,h)$ for two field orientations, along $z$ axis (green) and along $x$ axis (red),
for the set of parameters $J_1=J_3=1$ (ideal geometry), $J_2=3$, $\Delta=0.9,\,0.5,\,0$ (from top to bottom).
Exact-diagonalization data refer to finite periodic chains of $N=12$ sites.
Effective-model predictions (thin black curves) refer to thermodynamically large chains.}
\label{fig02}
\end{center}
\end{figure}

\begin{figure}
\begin{center}
\includegraphics[clip=on,width=75mm,angle=0]{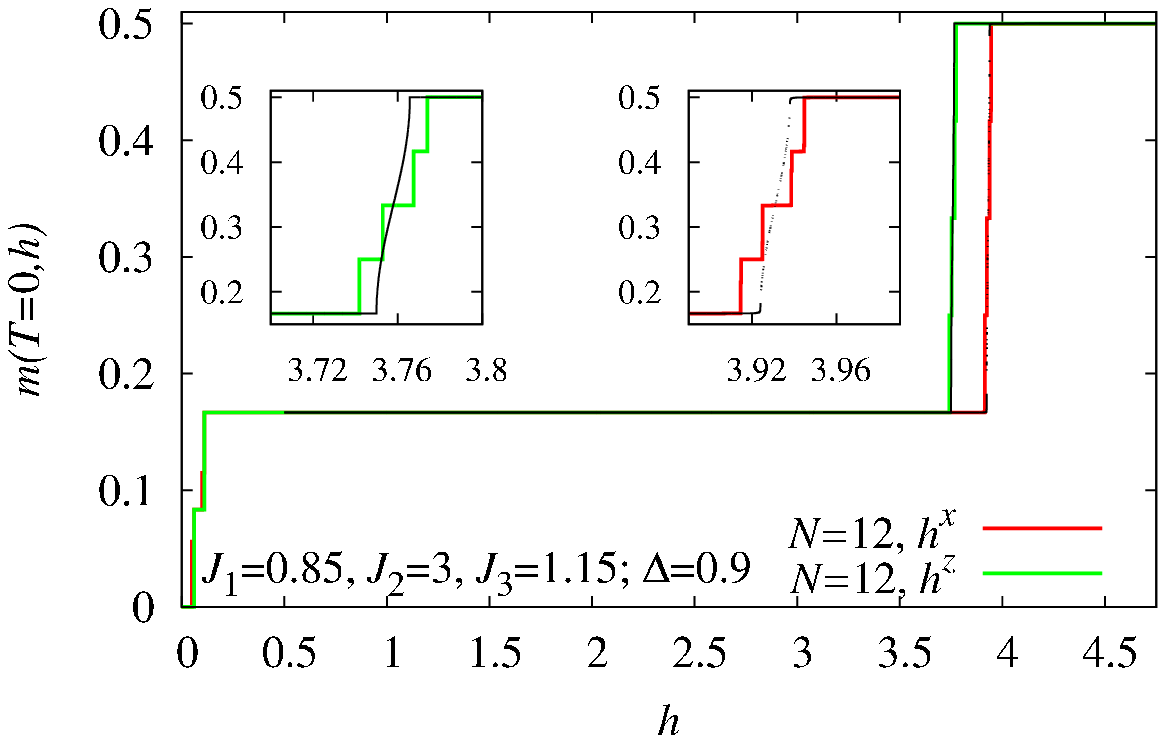}\\
\vspace{3mm}
\includegraphics[clip=on,width=75mm,angle=0]{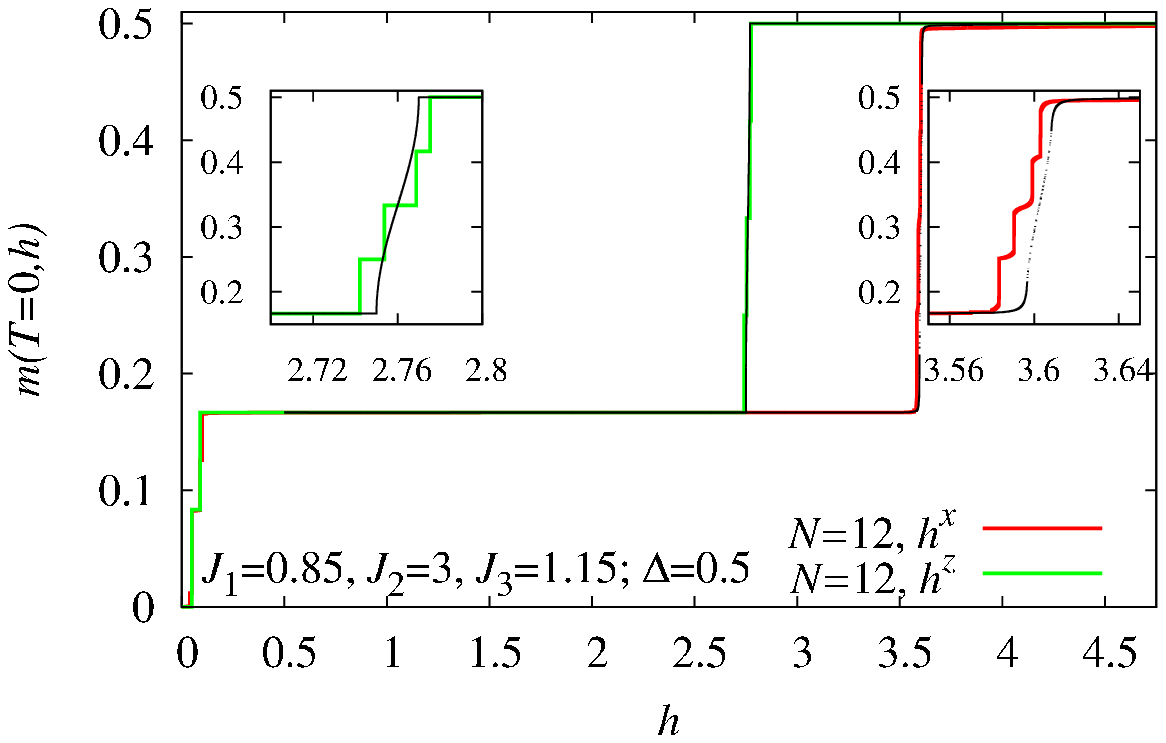}\\
\vspace{3mm}
\includegraphics[clip=on,width=75mm,angle=0]{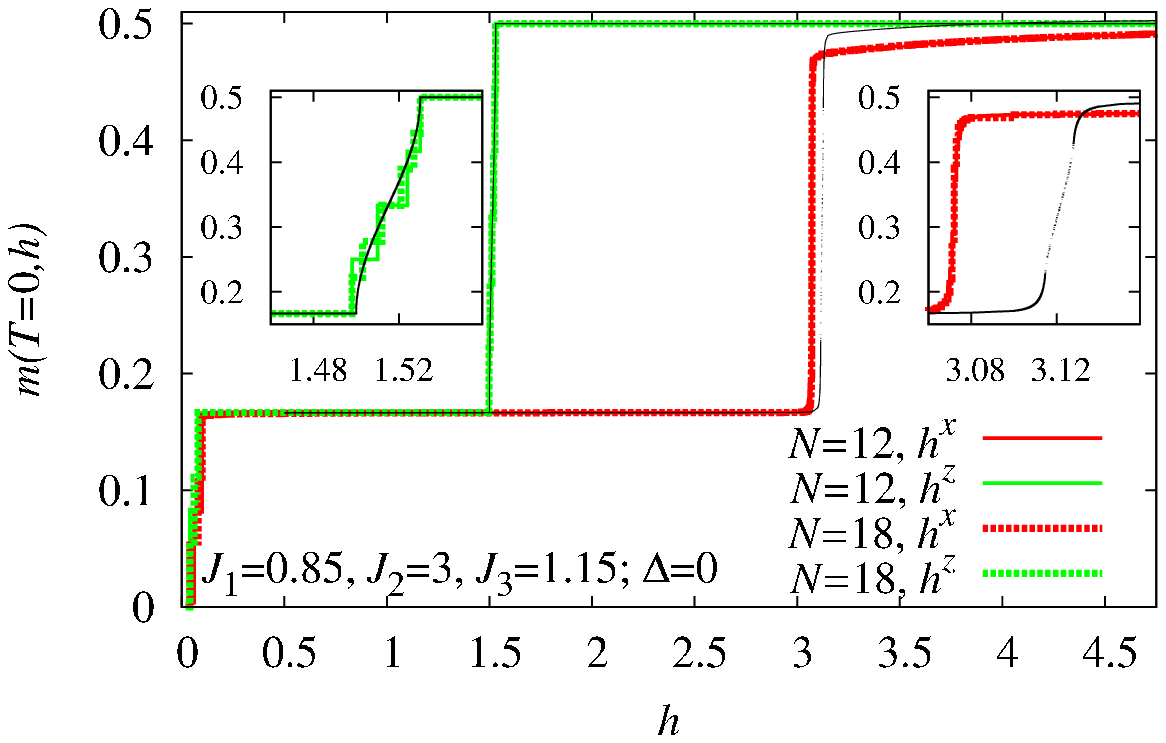}
\caption
{(Color online) The same as in Fig.~\ref{fig02}, however, for $J_1=0.85$, $J_3=1.15$ (distorted
geometry).
For the $XY$-limit ($\Delta=0$) we also report exact-diagonalization data for $N=18$ to
illustrate  finite-size effects.}
\label{fig03}
\end{center}
\end{figure}

\begin{figure}
\begin{center}
\includegraphics[clip=on,width=75mm,angle=0]{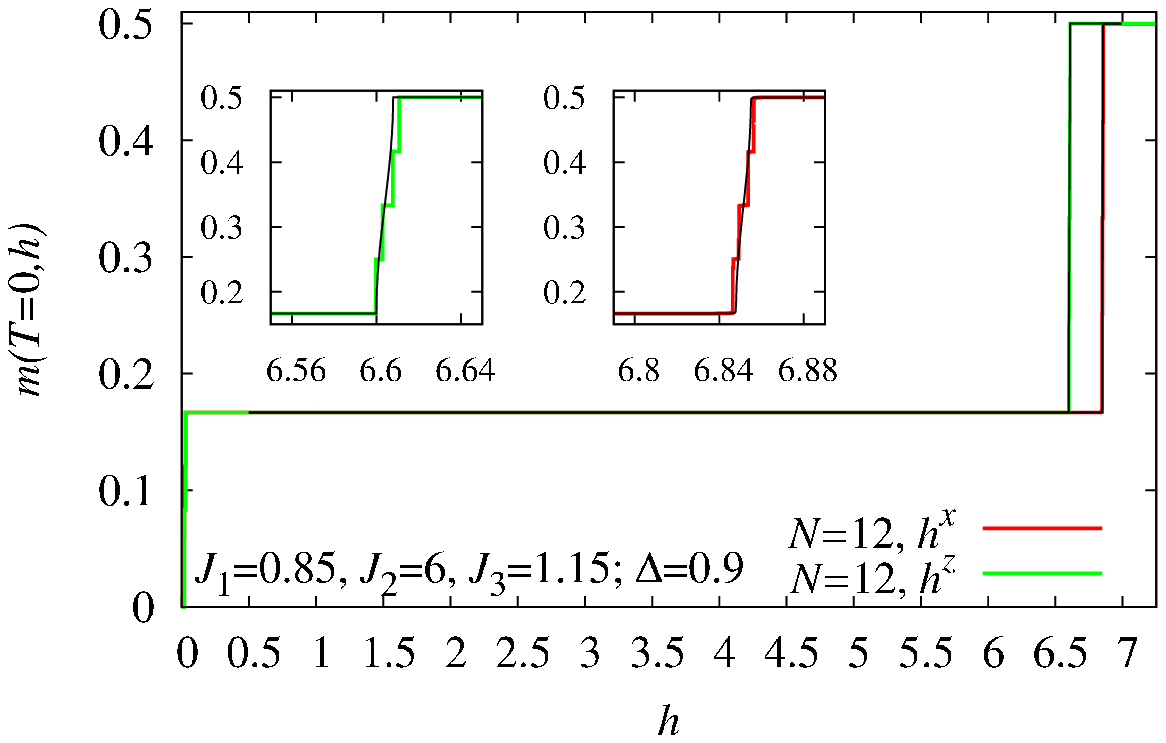}\\
\vspace{3mm}
\includegraphics[clip=on,width=75mm,angle=0]{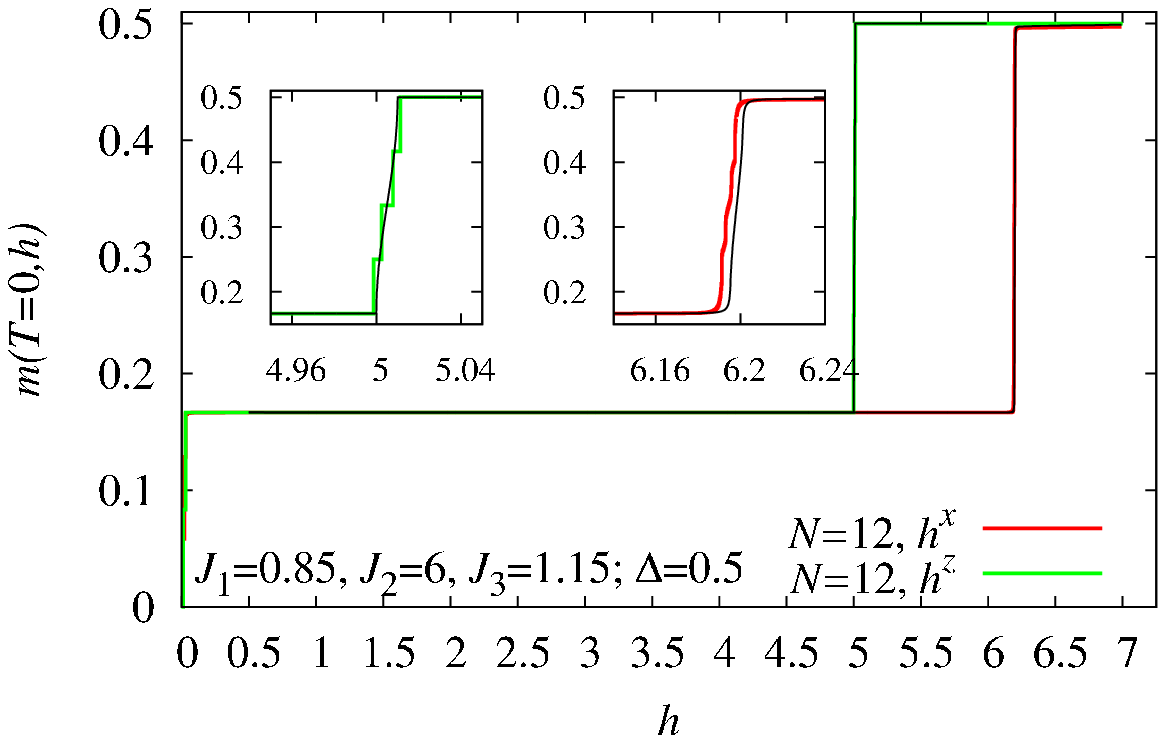}\\
\vspace{3mm}
\includegraphics[clip=on,width=75mm,angle=0]{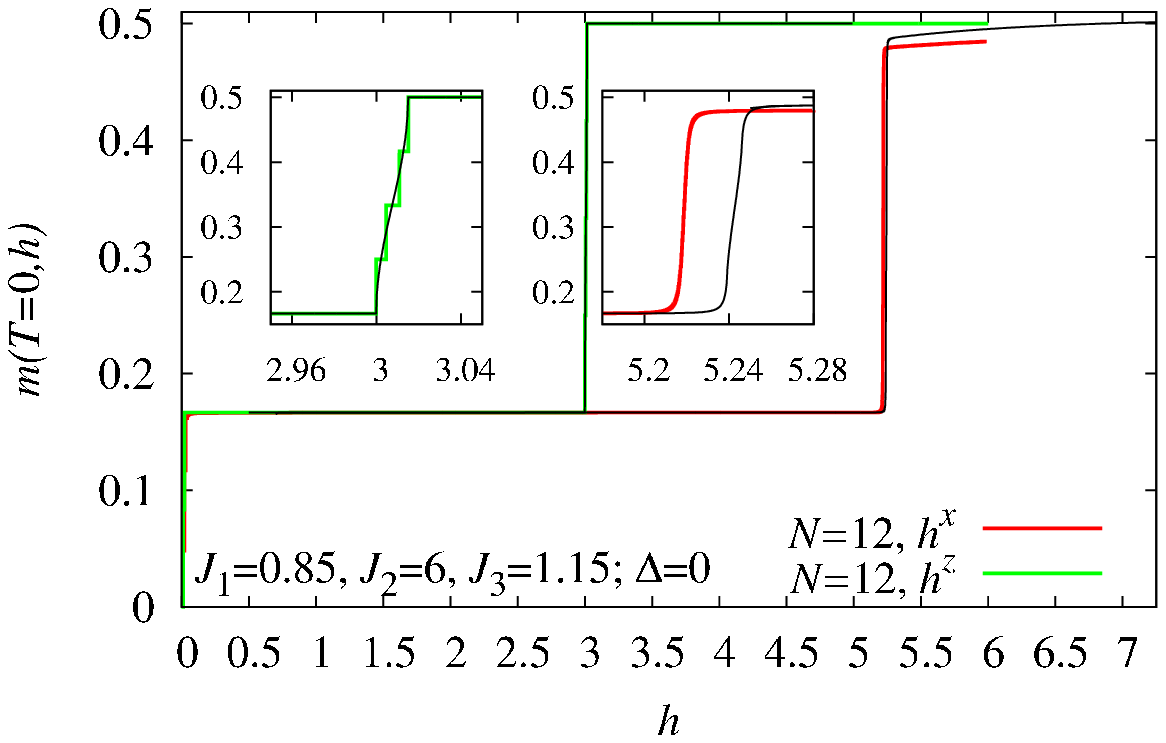}
\caption
{(Color online) The same as in Fig.~\ref{fig03} (distorted geometry), 
however, for a larger strength of the vertical dimer bond, $J_2=6$.}
\label{fig04}
\end{center}
\end{figure}

To test the quality of the elaborated low-energy effective theories 
we compare their predictions to finite-lattice exact-diagonalization data 
calculated for the initial frustrated quantum spin system.
We focus on the ground-state magnetization curves.
We consider systems of $N=12,\,18$ sites (i.e., ${\cal{N}}=4,\,6$ cells) imposing periodic boundary conditions
with
$J_1=J_3=1$ (ideal geometry), $J_1=0.85$, $J_3=1.15$ (distorted geometry), $J_2=3$, $J_2=6$, 
and $\Delta=0.9,\,0.5,\,0$, 
see Figs.~\ref{fig02}, \ref{fig03}, \ref{fig04}.

If $\Delta=1$ the magnetization curves for $z$- and $x$-aligned fields are identical.
However,
already for $\Delta=0.9$ the difference between the curves is obvious 
and it becomes more substantial for $\Delta=0.5$ and $\Delta=0$,
see Figs.~\ref{fig02}, \ref{fig03}, \ref{fig04}.
For the ideal-geometry case (Fig.~\ref{fig02}),
the ground-state magnetization curve has a jump at a (strong) characteristic field $h_*$.
For the $z$-aligned field, $h_*$ is the saturation field,
i.e., $m(T=0,h=h_*+0)=1/2$.
For $x$-aligned field, the saturation value of magnetization is achieved only in the limit $h\to\infty$,
however, $m(T=0,h=h_*+0)$ is very close to 1/2 even in the limit $\Delta=0$, 
see the corresponding panel in Fig.~\ref{fig02}.
[For $J_1=J_3=0$ we have $m(T=0,h=h_0+0)\approx 0.4809$ at $\Delta=0$.]
While the effective theory (\ref{02}) provides excellent description of the exact-diagonalization data,
the effective theory (\ref{09}) is less accurate especially in the limit $\Delta=0$.

Deviations from the ideal geometry lead to a smearing out of the magnetization jump.
For $z$-aligned field the effective theory (\ref{02}) provides only a qualitatively correct description of the magnetization profile 
as $\Delta$ is only slightly below 1 (see also Ref.~\citep{prb-fnt}),
however, its predictions agree excellently with exact-diagonalization data
for $\Delta \to 0$,
see the corresponding panel for $\Delta=0$ in Fig.~\ref{fig03}.
For $x$-aligned field the results for $\Delta=0.9$ as they follow from Eq.~(\ref{09}),
roughly speaking,
agree with exact-diagonalization data to the same degree as the ones 
which follow from Eq.~(\ref{02}) with exact-diagonalization data for $z$-aligned field,
cf. the corresponding curves in the panel for $\Delta=0.9$ in Fig.~\ref{fig03}.
However, when $\Delta$ becomes smaller the 
agreement between the effective theory given in Eq.~(\ref{09}) and the exact-diagonalization data becomes worse.
In particular, 
the effective theory (\ref{09}) overestimates the field at which the magnetization rapidly increases
towards the saturation value.
Moreover, 
the detailed profile of the magnetization curves also stronger deviates from the  exact-diagonalization data
(although this occurs at a very small scale),
see the right inset of panel for $\Delta=0$ in Fig.~\ref{fig03}.
It should be stressed that finite-size effects manifest themselves differently in the cases of $z$- and $x$-aligned fields if $\Delta$ approaches zero
and they are invisible for the $x$-magnetization in the scale used in the panel for $\Delta=0$ in Fig.~\ref{fig03}.

In Fig.~\ref{fig04} we show the results for nonideal geometry, however, with a larger value of $J_2$,
namely, $J_2=6$.
Comparing these data with the ones in Fig.~\ref{fig03} it is obvious 
that, as expected, the strong-coupling approach works much better for all values of $\Delta$, $\Delta=0.9,\,0.5,\,0$,
when $J_2$ increases.

Finally, 
we can estimate the characteristic field $h_*$ 
at which the ground-state magnetization jumps or almost jumps 
to the saturation value $1/2$ (or to a value close to $1/2$).
Within the effective theories an estimate for $h_*$ follows from the condition ${\sf{h}}(h_*)=0$.
For the $z$-aligned field we get
\begin{eqnarray}
\label{12}
h^z_*=\frac{1+\Delta}{2}J_2+\Delta J+\frac{(J_3-J_1)^2}{2(1+\Delta)J_2}.
\end{eqnarray}
For the $x$-aligned field and not too large $1-\Delta$ 
(e.g., for $\Delta=0.9,\, 0.5$ in Figs.~\ref{fig02}, \ref{fig03}, \ref{fig04})
we get approximately
\begin{eqnarray}
\label{13}
h^x_*\approx \sqrt{\frac{1+\Delta}{2}}J_2 + J
+\frac{(J_3-J_1)^2}{4\sqrt{\frac{1+\Delta}{2}}J_2}\Delta.
\end{eqnarray}
[To obtain this result we have set $a=1$ and $b=0$, see Eq.~(\ref{04}), which is a reasonable assumption for not too large $1-\Delta$].
These formulas give some reference fields above which the ground-state magnetization 
reaches (or becomes very close to) the saturation value.

\section{Conclusions}
\label{sec4}

In this paper we have analyzed how the anisotropy in the exchange interactions 
influences the low-temperature strong-field properties of a localized-magnon system.
As a paradigmatic model with relation to experimental findings we have 
examined spin-1/2 antiferromagnetic $XXZ$ Heisenberg model on a frustrated diamond-chain lattice 
in the presence of a (strong) magnetic field directed either along the $z$-axis or along the $x$-axis,
see Eq.~(\ref{00}) and Fig.~\ref{fig01}.
We have elaborated effective low-energy theories, see Eqs. (\ref{02}) and (\ref{09}),
which lead to exactly solvable spin models and provide a reasonable
description of thermodynamic quantities at low temperatures of the initial
frustrated quantum spin model.
In some detail we have analyzed the magnetization curves 
which may have some relevance to recent measurements on magnetic compounds related to the localized-magnon scenario.\citep{kikuchi,kikuchi-2,tanaka}

If the applied field is directed along the $z$-axis the localized-magnon picture holds:
The operator $S^z$ commutes with the Hamiltonian $H$
and therefore the eigenstates of $H$ can be classified according to the number of magnons $n=N/2-S^z$.
In the subspaces with $n=1,\ldots,N/3$ the lowest-energy magnons are localized, 
their contribution to the partition function dominates in the low-temperature strong-field regime 
and can be estimated along the lines of Ref.~\citep{locmag3}.
In the case of distorted flat-band geometry, the theory of Ref.~\citep{prb-fnt} is straightforwardly applicable.
The only changes conditioned by the exchange-interaction anisotropy in this case are quantitative.
We mention that the spin-1/2 isotropic $XY$ model in a $z$-aligned field with distorted diamond-chain geometry  
was treated approximately in Ref.~\citep{taras} by direct application of the Jordan-Wigner fermionization, 
however, not focused on the parameter regime considered in the present paper.

On the other hand,
if the field is directed along the $x$-axis the localized-magnon picture does not hold:
Since the Zeeman term does not commute with the Hamiltonian $H$ the number of localized
magnons is not a good quantum number any more.
However, the magnetization jump for the diamond chain survives as a prominent feature of localized-magnon systems.
This is a consequence of the fact 
that a state which contains local singlets (\ref{05}) on vertical bonds remains an eigenstate of the Hamiltonian (\ref{03}) even for
$0 \le \Delta<1$.
One may also expect that for small anisotropy $(1-\Delta)$ the localized-magnon theory is a good starting point for a description of the case.
In this study,
we have used alternatively the  strong-coupling treatment valid for small ratios $J_1/J_2$ and $J_3/J_2$, 
which does not require a small anisotropy $(1-\Delta)$.
Comparing the strong-coupling approach with exact-diagonalization data 
we have found that the strong-coupling approach works well up to quite large ratios  $J_1/J_2$ and $J_3/J_2$ 
if $(1-\Delta)$ is not too large.
In the large anisotropy limit $\Delta\to 0$ one needs smaller ratios  $J_1/J_2$ and $J_3/J_2$ 
to get reasonable agreement between strong-coupling and exact-diagonalization data.

The most prominent effect of the anisotropy in the exchange interaction $0\le \Delta <1$ 
is the shift of the steep increase (jump for ideal geometry, jump-like for nonideal  geometry) 
of the magnetization $m$ to saturation (or to nearly saturation) towards lower values of magnetic field
[see also Eqs.~(\ref{12}) and (\ref{13})].
The case of steep increase of $m$ to only nearly saturation is realized, if the field is aligned along the $x$-axis. 
Then the magnetization becomes fully saturated only if the field is infinitely large.

For future  studies it is of interest
to apply the strong-coupling approach to other frustrated quantum antiferromagnets
considered in Ref.~\citep{prb-fnt}, 
in particular, to two-dimensional systems.
Moreover, 
an explanation of the experimental data for the spin dimer magnet Ba$_2$CoSi$_2$O$_6$Cl$_2$\citep{tanaka} would be a challenging
project.

\section*{Acknowledgments}

Exact-diagonalization calculations were performed using ALPS
package\citep{alps} and J.~Schulenburg's {\it spinpack}.\citep{spinpack}
The present study was supported by the DFG (project RI615/21-1).
O.~D. acknowledges the kind hospitality of the University of Magdeburg in
April-June of 2014.
O.~D. would like to thank the Abdus Salam International Centre for
Theoretical Physics (Trieste, Italy)
for partial support of these studies through the Senior Associate award.





\end{document}